\documentclass[preprint,authoryear,11pt]{elsarticle}
\usepackage[margin=0.80in]{geometry}
\usepackage{float}
\usepackage{booktabs}
\usepackage{array,ragged2e}
\usepackage{wrapfig}
\usepackage[utf8x]{inputenc}
\usepackage{graphics}
\usepackage{graphicx}

\usepackage{caption}
\usepackage{subcaption}

\usepackage{caption}
\usepackage{adjustbox}
\usepackage{wrapfig}
\usepackage{array,ragged2e}
\usepackage[table]{xcolor}
\usepackage{amssymb,amstext,amsmath}
\usepackage{float}
\usepackage{multicol}
\usepackage{tabularx}
\usepackage{multirow}

    \makeatletter
    \def\ps@pprintTitle{%
       \let\@oddhead\@empty
       \let\@evenhead\@empty
       \let\@oddfoot\@empty
       \let\@evenfoot\@oddfoot
    }
    \makeatother

\usepackage[nodisplayskipstretch]{setspace}
\begin{document}

\begin{frontmatter}


\title{Evaluation of Sustainable Green Materials: Pinecone in Permeable Adsorptive Barriers for Remediation of Groundwater Contaminated by $Pb^{2+}$ and Methylene Blue}
\author[rvt,fn1]{Samuel Darko\corref{cor1}}
\author[rvt]{Gurcan Comert}
\author[focal2]{Noelle A Mware}
\author[focal4]{Faith Kibuye}
\address[rvt]{Computer Science, Physics, and Engineering Department, Benedict College, Columbia, SC 29204 USA}
\address[focal2]{Department of Civil and Environmental Engineering, University of Nebraska-Lincoln, Lincoln, NE 68588 USA}
\address[focal4]{Southern Nevada Water Authority, Las Vegas, NV 89153 USA}
\fntext[fn1]{Corresponding author, samuel.darko@benedict.edu}
\begin{abstract}
We report herein, the potential of raw pinecone powder (PCP) and pinecone biochar (PCBC) as alternatives to activated carbon used in Permeable Adsorptive Barriers (PABs) for the in situ remediation of polluted groundwater. A constructed lab-scale unconfined aquifer ($38\times30\times17$ $cm$) fitted with PCP and PCBC PABs ($21\times3\times20$ $cm$), was evaluated for the removal of $Pb^{2+}$ ions in a continuous flow setup. Results indicate that after $3600$ minutes, PCP was able to reduce $Pb^{2+}$ ions from a Co=$50$ $mg/L$ to $7.94$ $mg/L$ for the first run and $19.4$ $mg/L$ for a second run, respectively. Comparatively, PCBC reached $6.5$ $mg/L$ for the first run and $8.94$ for the second run. It was confirmed that adsorption was best described by the first-order kinetic model with $R^2$ values above $0.95$. Maximum adsorption capacity values were found to be $1.00$, $0.63$, $1.08$, and $0.85$ mg/g for each scenario respectively. In addition, nonlinear regression models of exponential and Gaussian Processes are fit to explain remediation by time for $Pb^{2+}$ and Methylene Blue. Gaussian Processes are able to better explain the variation of pollution removal compared to simpler exponential models. When regressed against true removal percentages all models are able to provide $R^2>0.99$.  
\end{abstract}
\begin{keyword}
Permeable Adsorptive Barriers, Pinecone Char, $Pb^{2+}$, Methylene Blue, Gaussian Process Regression, green materials
\end{keyword}
\end{frontmatter}
\section{Problem Definition}
\label{sctprb}
Groundwater contamination by heavy metals such as $Pb^{2+}$ is a huge environmental concern because groundwater is a major source of water for day today application. The presence of Pb2+ in groundwater is a direct hazard to human beings causing severe neurological complications as well as organ and tissue damage (\cite{mohammad2010metal}). Pb2+ pollution in groundwater can be related to various anthropogenic sources including: landfill waste disposal (\cite{oyeku2010heavy}), industrial and agricultural activities (\cite{amin2011groundwater}). Groundwater pollution could also occurs as a result of water exchange from contaminated surface waters or leaching from contaminated soils through rainfall infiltration (\cite{di2008groundwater}).  
In situ remediation of contaminated groundwater sites presents more advantages in relation to ex situ pump and treat technologies including: feasibility and effectiveness of the method (\cite{di2008groundwater}); its versatile and simple configuration (\cite{erto2014permeable}); and low operation costs and energy requirements (\cite{park2002lab}). In situ treatment makes use of Permeable Reactive Barriers (PRB) made of zero valent iron (Fe0) as the reactive material (\cite{di2008groundwater,erto2014permeable,park2002lab,blowes2000treatment,kamolpornwijit2003preferential,simon2002groundwater,komnitsas2006inorganic,di2006groundwater,keith2005effect}). The barriers are inserted underground to intercept the natural flow of contaminated groundwater allowing the water to pass through the barrier (\cite{erto2014permeable}) and is passively treated through redox and precipitation reactions (\cite{di2008groundwater}). The main shortcoming of the Fe0 barriers is the accumulation of precipitates that may limit barrier longevity by reducing porosity and conductivity (\cite{blowes2000treatment}) and preferential flow paths in the barrier may form leading to a reduction of the contact time of the contaminated water with the reactive material (\cite{kamolpornwijit2003preferential}).
Researchers have therefore investigated the possible application of adsorbents as alternative materials to $Fe^0$ in Permeable Adsorptive Barriers (PAB). The main adsorbent applied in PABs is activated carbon (\cite{di2008groundwater,erto2014permeable}). Due to the environmental impact and costly nature of activated carbon, other adsorbents such as zeolites (\cite{park2002lab}) have been investigated for the application in PABs.
\begin{figure}[h] 
\centering
\includegraphics[scale=.25]{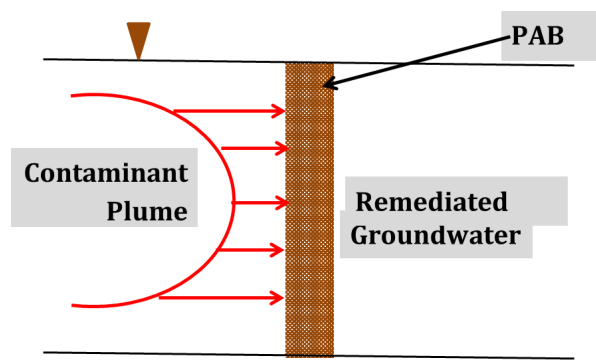}
\caption{Permeable adsorptive barrier installed along the groundwater flow of contaminated plume}
\label{fig_pab}       
\end{figure}
 
In this study we investigated the possible application of cheaply acquired and environmentally sustainable raw pinecone powder (PCP) and pinecone biochar (PCBC) as adsorbents in PABs for in situ remediation of groundwater contaminated by $Pb^{2+}$ in a lab-scale constructed unconfined aquifer. 
\section{Materials and Methods}
\label{sctmm}
\subsection{Adsorbent Fabrication}
\label{sctfab}
Sun-dried pinecones were collected from the backyard of the principal investigator in Columbia, South Carolina, USA.  The cones were scaled and washed with deionized water to remove all traces of impurities. Washed scales were subsequently oven dried at $50$ $^{\circ}C$ for 1 hour then ground into finer particles using a planetary ball mill to obtain raw pinecone powder (PCP).
\begin{figure}[h] 
\centering
\includegraphics[scale=.21]{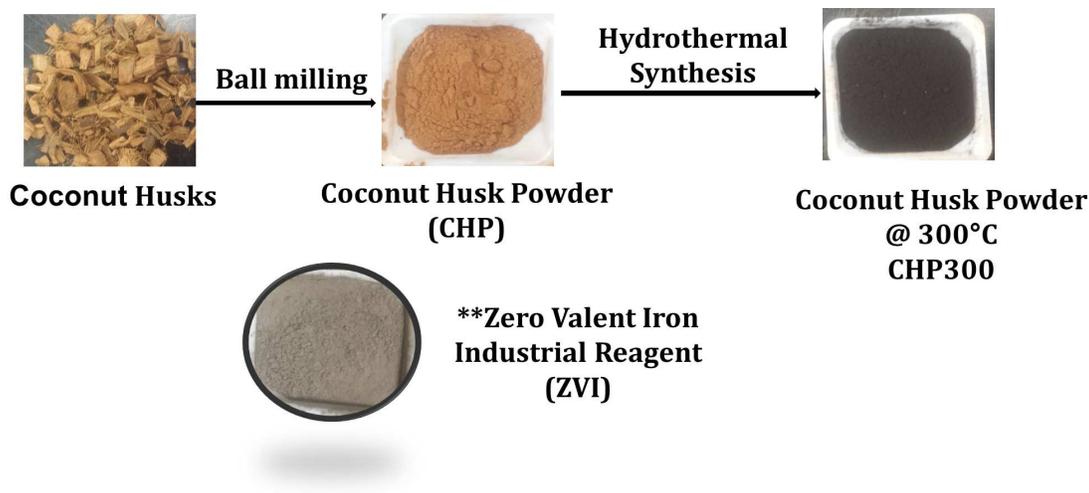}
\caption{PCP and PCBC derived from grinding pinecone and hydrothermal carbonization  of PCP at $300$ $^{\circ}C$, respectively}
\label{fig_fab}       
\end{figure}

Portions of the acquired powder was converted into pinecone biochar through hydrothermal carbonization (HTC) performed as follows: $20$ grams of the PCP and $40$ $ml$ of deionized water were packed into custom $200$ $ml$ HTC reactor and oven heated at $300$ $^{\circ}C$ for $4$ hours to obtain  respective pinecone biochars (PCBC).

The physico-chemical properties of the adsorbents were determined through material characterization using a PerkinElmer Fourier transform infrared spectrometer (FTIR), a Zeiss Ultra Plus Field Emission Scanning Electron Microscope (FESEM), a Shimadzu X-ray diffraction (XRD), and Micrometrics ASAP 2020 Brunner-Emmett-Teller (BET) analyzer. 
\subsection{Experimental Setup}
\label{sctexp}
\begin{figure}[h] 
\centering
\includegraphics[scale=.19]{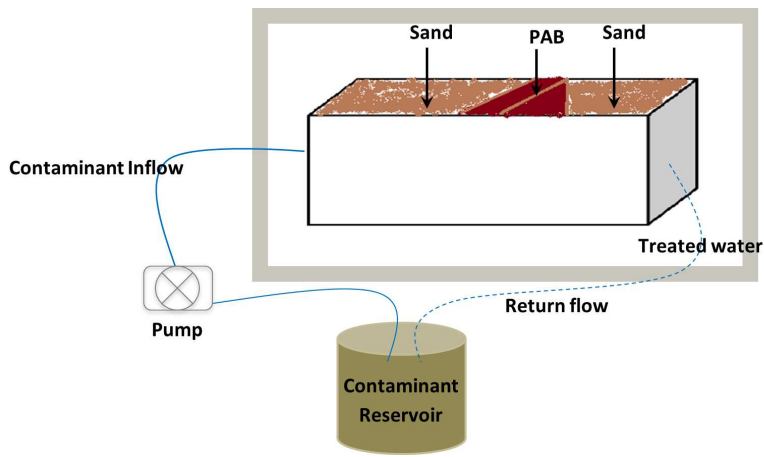}
\caption{Experimental design- flow diagram  of the contaminated groundwater through the PAB}
\label{fig_dflow}       
\end{figure}

\begin{figure}[h] 
\centering
\includegraphics[scale=.19]{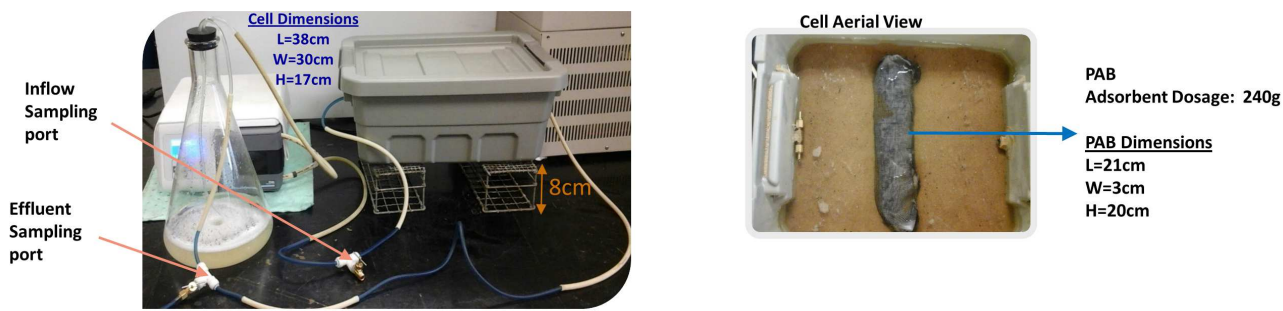}
\caption{Experimental set-up for  laboratory simulation of PAB  bound within sand in the cell.}
\label{fig_edes}       
\end{figure}

A lab-scale constructed cell emulating an unconfined aquifer (38x30x17cm) was packed with clean sand of known particle size distribution (Fig~\ref{fig_dflow}). The cell was then centrally fitted with PCP and PCBC PABs (21x3x20cm). A peristaltic pump set at a constant pumping rate of 70ml/min was used to continuously pump 6 $L$ of 50 $mg/L$ of $Pb^{2+}$ from a contaminant reservoir via an influent tubing into the cell (Fig~\ref{fig_edes}). The total contact time between the contaminant and each PABs was $3600$ minutes. Samples were collected using an effluent sampling port at 10 minute interval within the first 60 minutes and subsequently after every 60 minutes. The same experimental setup was repeated with reused PABs to investigate the residence time of the PABs. Collected samples were analyzed for final concentrations using an air/acetylene Agilent Atomic Absorption (AA) Spectrophotometer.

\section{Results}
\label{sctres}
\subsection{Adsorbent Characterization}
\label{sctchar}
\begin{figure}[h!]
\centering
\begin{subfigure}{.46\textwidth}
\centering
  \includegraphics[width=1.0\linewidth]{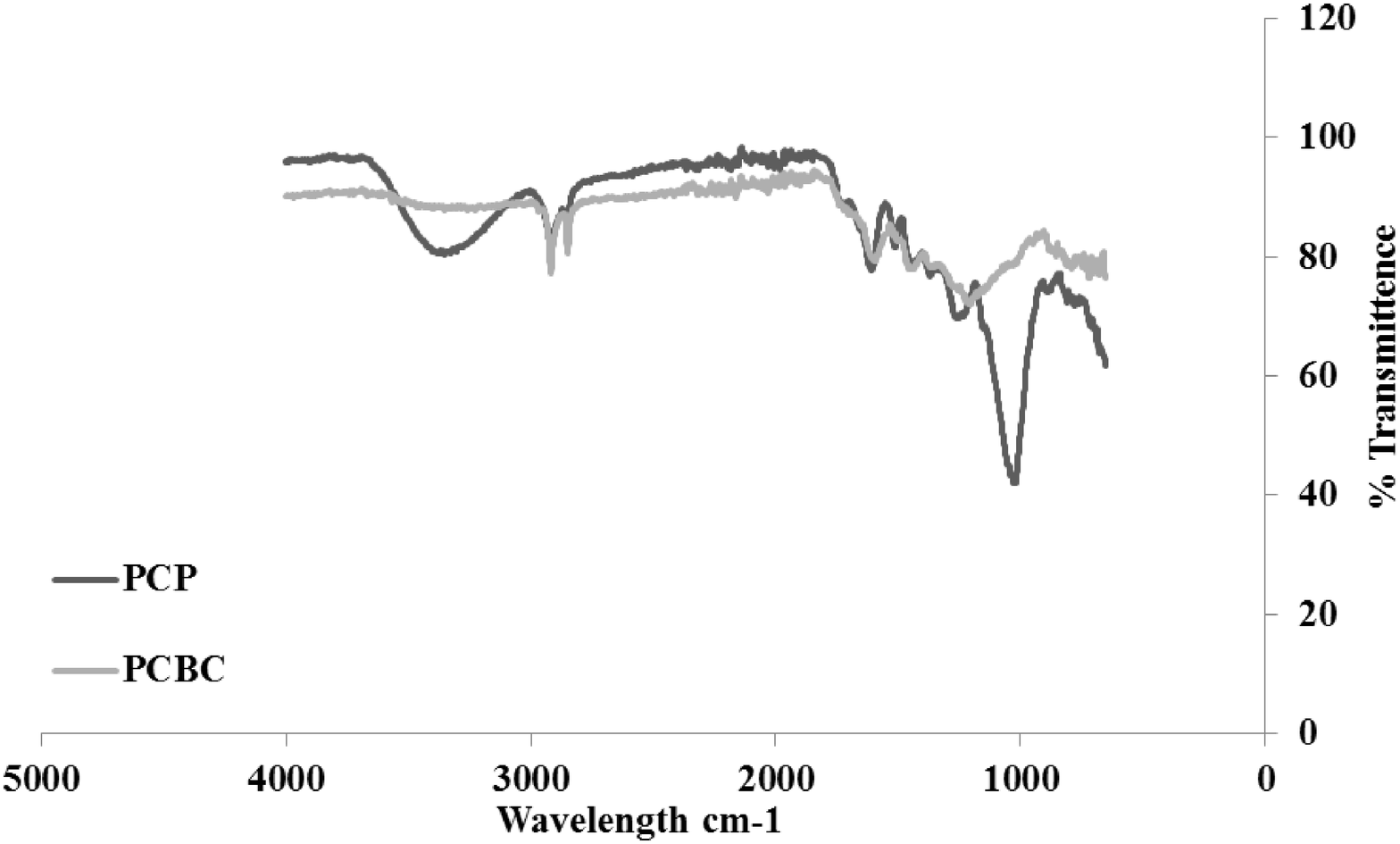}
  \caption{Physicochemical characterization of PCP and PCBC}
  \label{fig_ftirr}
\end{subfigure}%
\begin{subfigure}{.51\textwidth}
 \centering
  \includegraphics[width=1.0\linewidth]{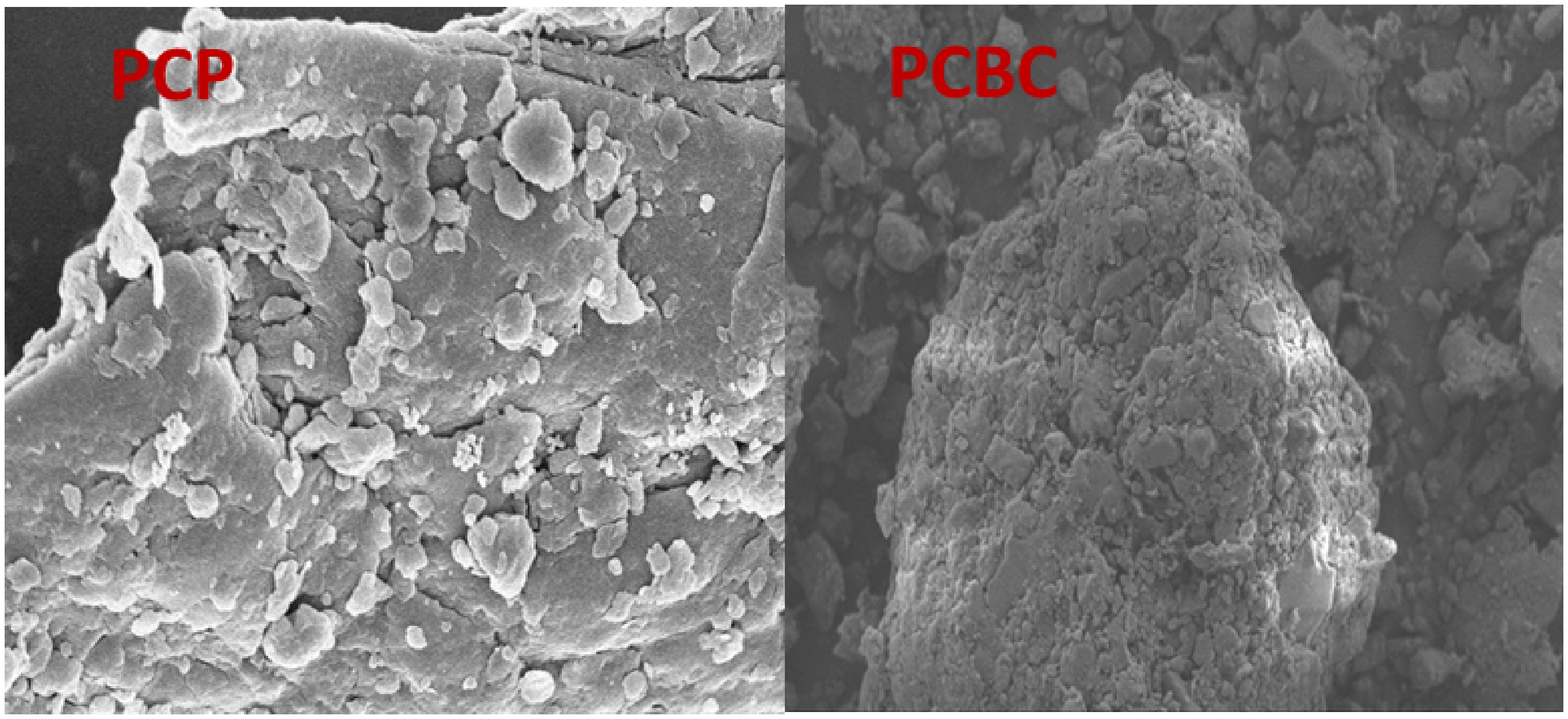}
  \caption{scanning electron micrographs images}
  \label{fig_ftir2}
\end{subfigure}
\caption{Physicochemical characterization of PCP and PCBC -  surface area and pore size  by BET analysis, crystallinity by XRD , functional groups determined by FTIR analysis and  scanning electron micrographs images}
\label{fig_ftir}
\end{figure}

The FTIR spectra were measured within a scan range of 4000-650 $cm^{-1}$ wave number. According to the spectra shown in Fig.~\ref{fig_ftirr} PCP and PCBC have shown the presence of essential functional groups on their surfaces. The infrared spectra bands observed at $3357.53$ $cm^{-1}$ and $2918.61$ $cm^{-1}$ represent the carboxylic acid group $O-H$ and aliphatic $C-H$ stretches. The peak at $1613.69$ $cm^{-1}$ corresponds to a conjugated $C=C$ stretch while $1020.08$ $cm^{-1}$ peak represented a $C-O$ stretch (\cite{ofomaja2009removal}). It is observed that the intensity of the functional group bands, stretches and peaks were lower in PCBC compared to PCP. These variations in intensity indicate that diverse thermal decompositions transpired throughout the HTC process resulting in loss of various functional groups on the surface of PCBC (\cite{al1997sorption}). 

The SEM images shown in Fig.~\ref{fig_ftir2} indicates the surface morphology of the adsorbents. The presences of pores on pinecone materials indicate that they present good characteristics to be employed as adsorbents for metal ions uptake (\cite{al1997sorption}). PCP consist of a smooth flat multilayer surface with tiny pores while PCBC depicted a coarser and rougher surface and larger pores. This change in the microstructure of PCBC can be attributed to the release of volatile gas contained in soften biomass matrix during the HTC processing (\cite{do2012biomass}). 
The XRD analysis revealed that pine cone materials are highly amorphous indicating that they can be used as good adsorbents for heavy metals (\cite{wuana2011heavy}). PCBC had a slightly higher percent amorphous zone of $88.67\%$ than that of PCP of $87.75\%$. 

The BET surface areas analysis of the adsorbents showed that PCP had a higher surface area of $4.43$ $m^2/g$ than that of PCBC of $3.07$ $m^2/g$. The average pore sizes were $198.64$ $Å$ and $189.90$ $Å$ for PCBC and PCP respectively. 
\subsection{$Pb^{2+}$ Adsorption Results}
\label{sctpb2}
\begin{figure}[h] 
\centering
\includegraphics[scale=.3]{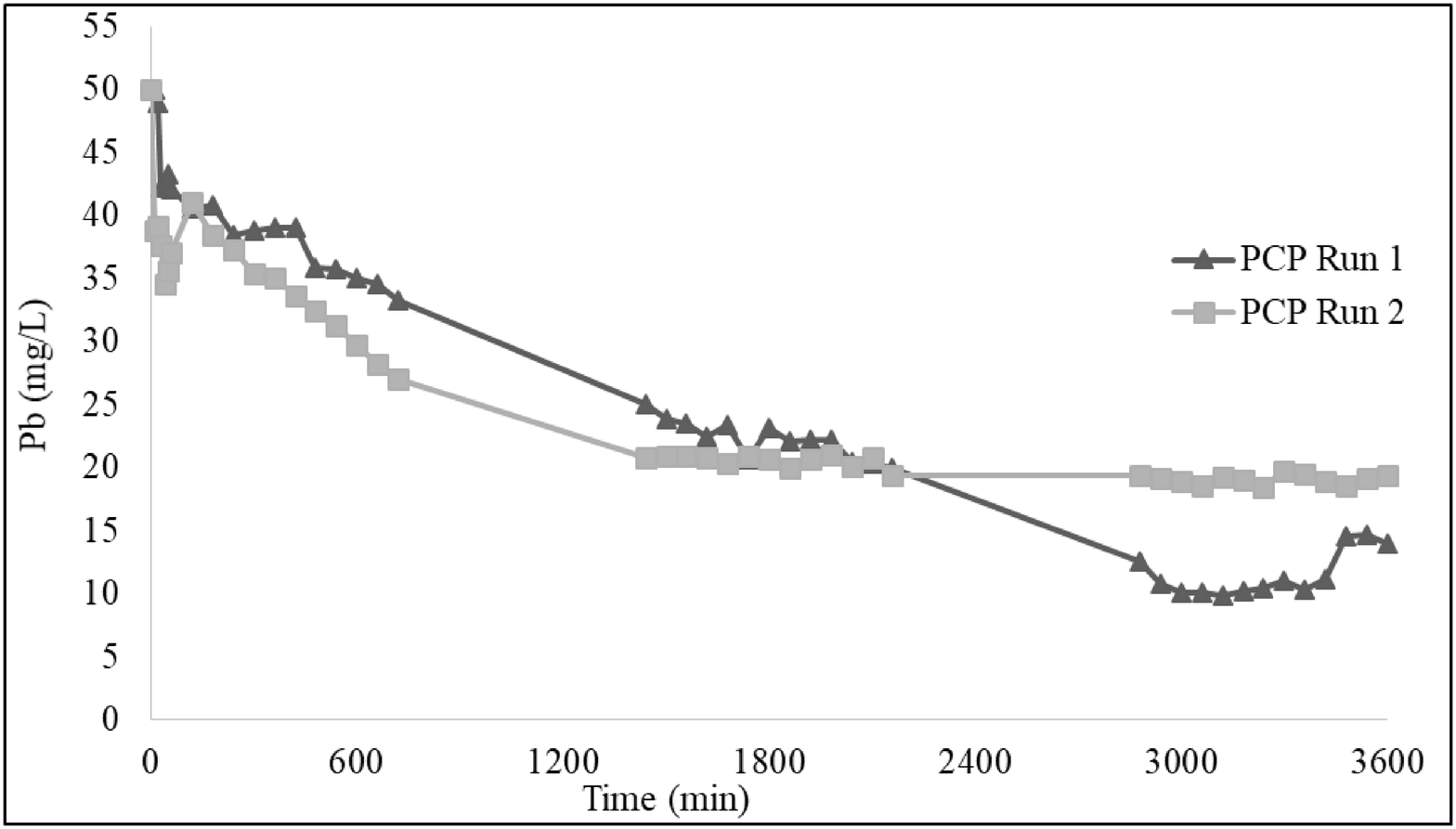}
\caption{$Pb^{2+}$ using PCP}
\label{fig_3}       
\end{figure}

\begin{figure}[h] 
\centering
\includegraphics[scale=.3]{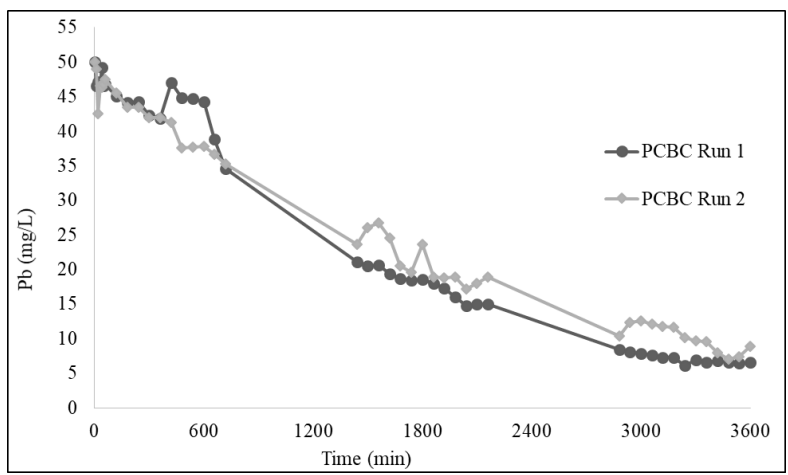}
\caption{$Pb^{2+}$ using PCBC}
\label{fig_4}       
\end{figure}

The effect of contact time and type of adsorbent barrier on the removal of $Pb^{2+}$ from a simulated unconfined aquifer was evaluated. The efficiency of the PCP and PCBC was further investigated by spiking the system with fresh $50$ $mg/L$ of $Pb^{2+}$. Obtained results were plotted final concentrations versus time as shown in Figs.~\ref{fig_3} and Fig.~\ref{fig_4}. PCP PAB, Fig.~\ref{fig_3}, resulted in a decrease in final concentration with increasing contact time. The lowest concentration attained with PCP was $9.86$ $mg/L$ at time $3120$ minutes. Beyond this time a steady rise in final concentration to a final concentration of $14$ $mg/L$ at time $3600$ minutes was observed, an indication possible instances of desorption. The second run using PCP PAB resulted in a much lower adsorption efficiency. Similar decrease in final concentration with time was observed, however an equilibrium was attained at time $1440$ minutes with a final concentration of $20.8$ $mg/L$.

The adsorption trend depicted by PCP can be attributed to its characteristics. PCP depicted a higher BET surface area and a highly amorphous structure that could explain its high adsorption efficiency in the first run. The lower adsorption efficiency in the second run can be attributed to the high intensity of acidic functional group $O-H$ on PCP. The $O-H$ group could have formed intermolecular bonds with H2O molecules from the first run causing the surface of PCP to be covered $H_2O$ molecules hence minimal adsorption of $Pb^{2+}$. 
PCBC PAB, Fig.~\ref{fig_3}, depicted a higher adsorption efficiency for $Pb^{2+}$ compared to PCP. A decrease in final concentration with contact time was observed. The adsorption with PCBC attained the lowest concentration of $6.53$ $mg/L$ at time $3600$ with no observable equilibrium. The second run with PCBC also depicted similarly high adsorption efficiency with final concentration of $8.94$ $mg/L$ at time $3600$ minutes.

The high adsorption efficiency for PCBC can also be related to its properties. PCBC depicted a higher amorphous structure that PCP. This property allows it to have more intermolecular space to for $Pb^{2+}$ to be adsorbed. It was also reported to have a high average pore size that can be attributed to its high adsorption efficiencies in both first and second runs. PCBC further depicted minimal $O-H$ functional groups, this could have resulted in minimal competition of its adsorption sites, allowing $Pb^{2+}$ to be adsorbed more efficiently.
\subsection{Pseudo First-Order Kinetic Model}
\label{sctkm}
\begin{figure}[h] 
\centering
\includegraphics[scale=.75]{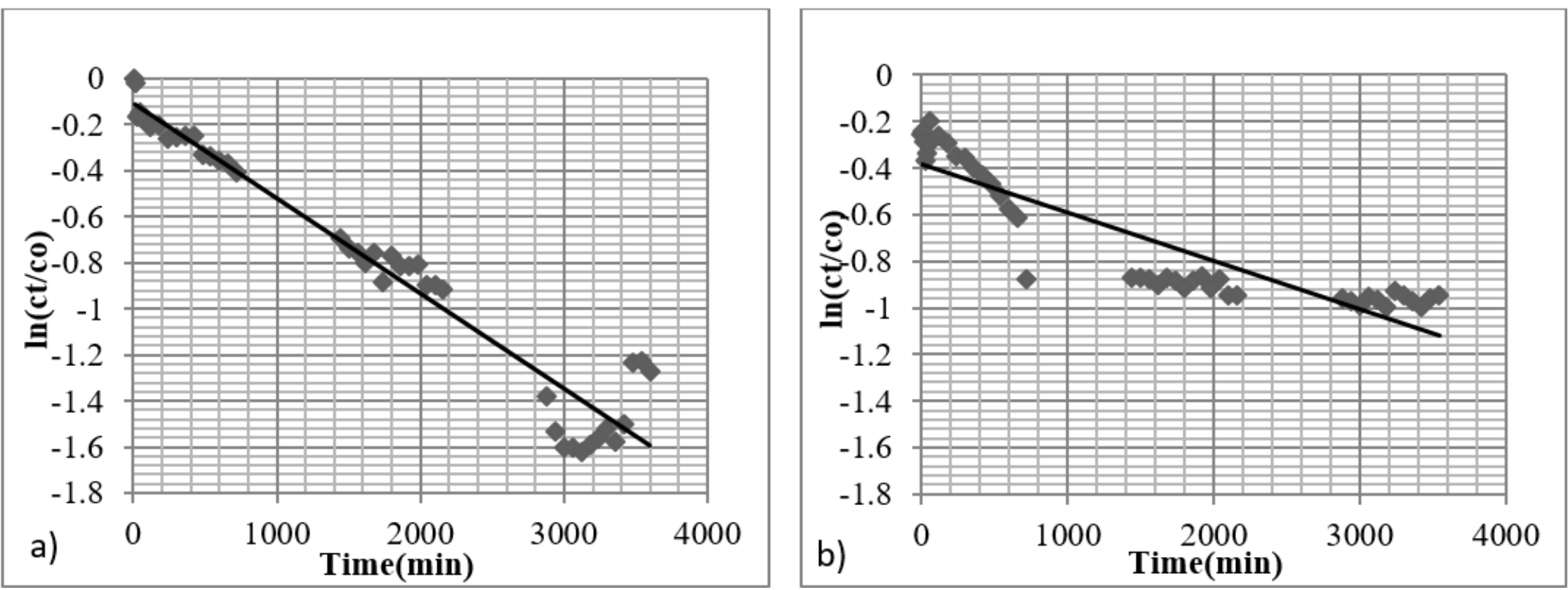}
\caption{First-order-kinetic plots of PCP a) Run 1 and b) Run 2}
\label{fig_5}       
\end{figure}

\begin{figure}[h] 
\centering
\includegraphics[scale=.75]{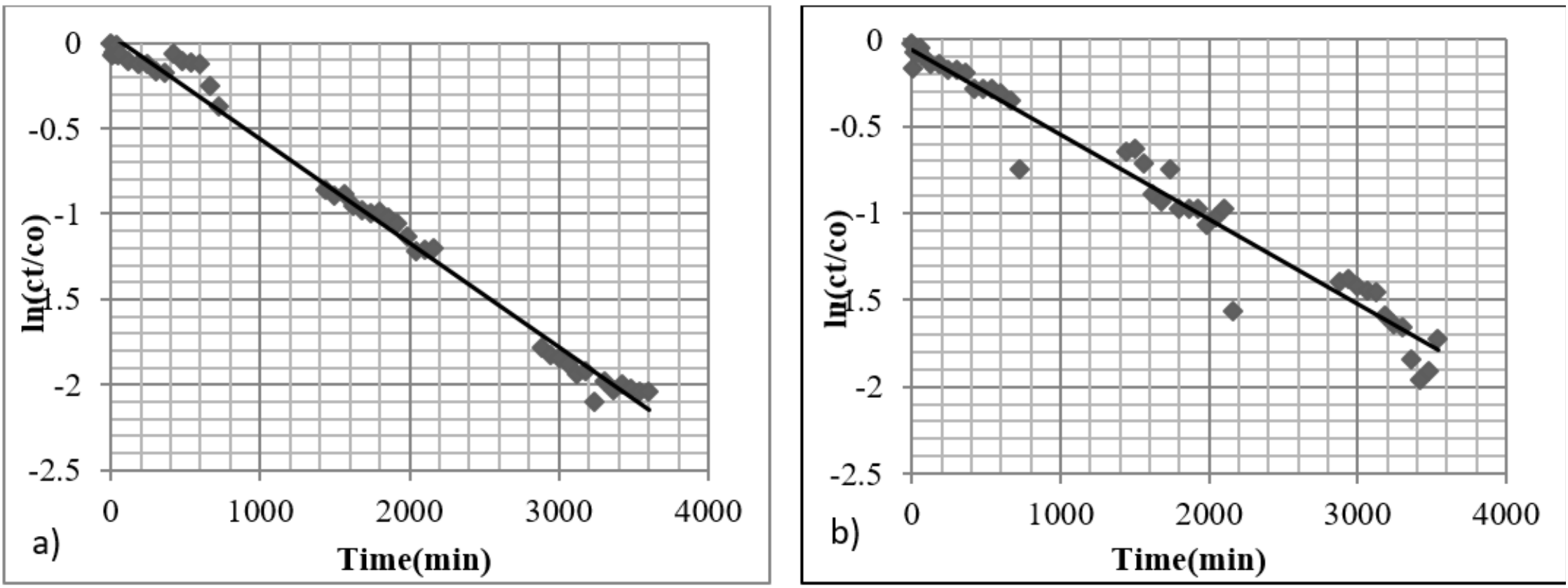}
\caption{First-order-kinetic plots of PCBC a) Run 1 and b) Run 2}
\label{fig_6}       
\end{figure}

Kinetic analysis of adsorption results is necessary as it provides insight on the rates of adsorption mechanism which are used primarily in the modeling and design process (\cite{jeyakumar2014adsorption}) and is important in analyzing the efficiency of the process (\cite{bulut2007removal}).
The experimental data were fitted in the linear form of the pseudo-first-order rate Eq.~(\ref{eqn_km}) given as: 
\begin{equation}
ln[c_t] =−k_t + ln[c_o]. 
\label{eqn_km}  
\end{equation}                                                                                                
Where [$c_t$] and [$c_o$] are final and initial concentrations of $Pb^{2+}$ respectively and $k$ ($min^{-1}$) is the pseudo-first-order adsorption rate constant. The linear plots of $log(c_t/c_o)$ versus $t$ (minutes) drawn for the pseudo-first-order model are shown in Figs.~\ref{fig_5} and \ref{fig_6}. The slopes of the plots of the experimental data and the intercepts of were used to determine the rate constant $k$ and the corresponding regression correlation coefficient $R^2$ values. The results are given in Table~\ref{tab_reg}.   

The correlation coefficients for the second-order kinetics model ($R^2$) are greater $0.9$, indicating the applicability of this kinetics equation and the first-order nature of adsorption process of $Pb^{2+}$ onto PCB and PCBC PABs.

\begin{table}[!htb]
\centering
\caption{Rate constant $k$ and $R^2$ values from first-order-kinetic model}
\label{tab_reg} 
\scalebox{0.95}{    
\begin{tabular}{c c c}
\hline\noalign{\smallskip}
 Adsorptive Barrier & Rate Constant ($k$) & $R^2$ \\
\noalign{\smallskip}\hline\noalign{\smallskip}
PCP $1^{st}$ Run & $-0.0004$ & 0.9439\\
PCP $2^{nd}$ Run &  $-0.0002$ & 0.8081\\
PCBC $1^{st}$ Run & $-0.0006$ & 0.9894\\
PCBC $2^{nd}$ Run &  $-0.0005$ & 0.9634\\
\noalign{\smallskip}\hline\noalign{\smallskip}
\end{tabular}
}
\end{table}

\subsection{Nonlinear Regression Models}
\label{sctm}
\begin{figure}[h] 
\centering
\includegraphics[scale=.6]{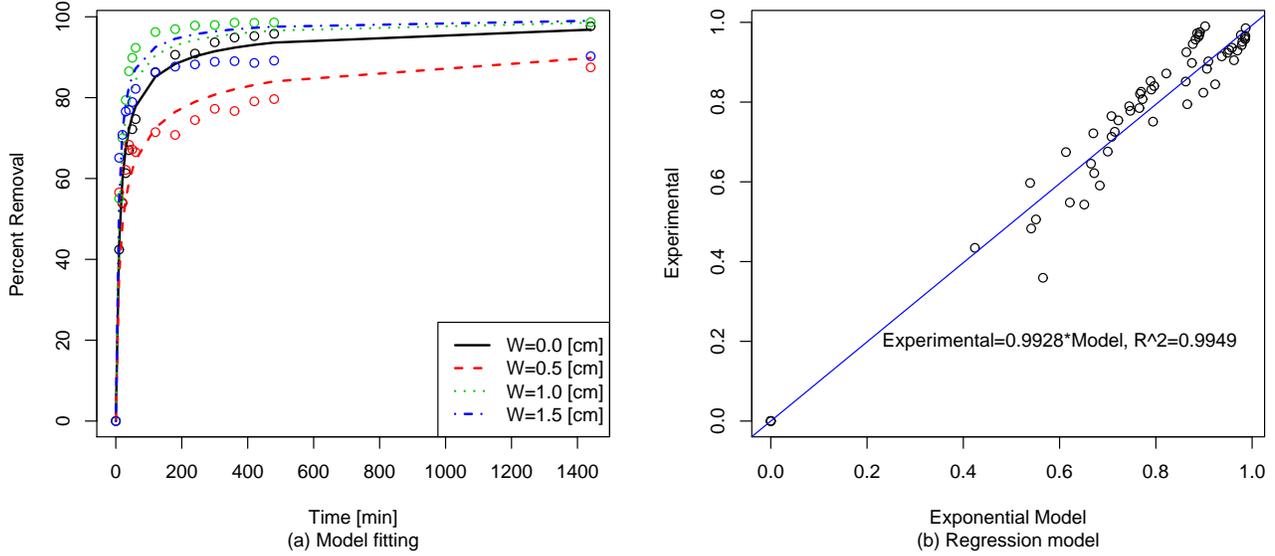}
\caption{Experimental versus exponential model of the removal of Methylene Blue.}
\label{fig_mat2}       
\end{figure}

Experimental design- flow diagram  of the contaminated groundwater through the PAB. In this subsection, results from models fitted for percent removal with respect to time ($t$) $min$ and thickness ($W$) $cm$ are presented. Out of several models fitted, based on the root mean squared error, minimum and simplest model is presented here. Within several nonlinear regression models: hyperbolic tangent, logistic, exponential (see Eq.~(\ref{eqn_exp})), and Gaussian Processes, latter two provided better results. Model parameters are optimized by gradient descent. GPs kernel coefficients are considered hyperparameters. The time was transformed as $t=\frac{ln(t')}{max(ln(t'))}$. 
\begin{equation}
C=1-e^{-t(a+(b+W))}-ta(b+W)e^{-t(a+(b+W)} 
\label{eqn_exp}  
\end{equation}

where, for $Pb^{2+}$ $a=3.315$ and $b=0.829$, and $W$ denotes the thickness. For Methylene Blue, $a=2.068$ and $b=3.486$.  
\begin{figure}[h] 
\centering
\includegraphics[scale=.6]{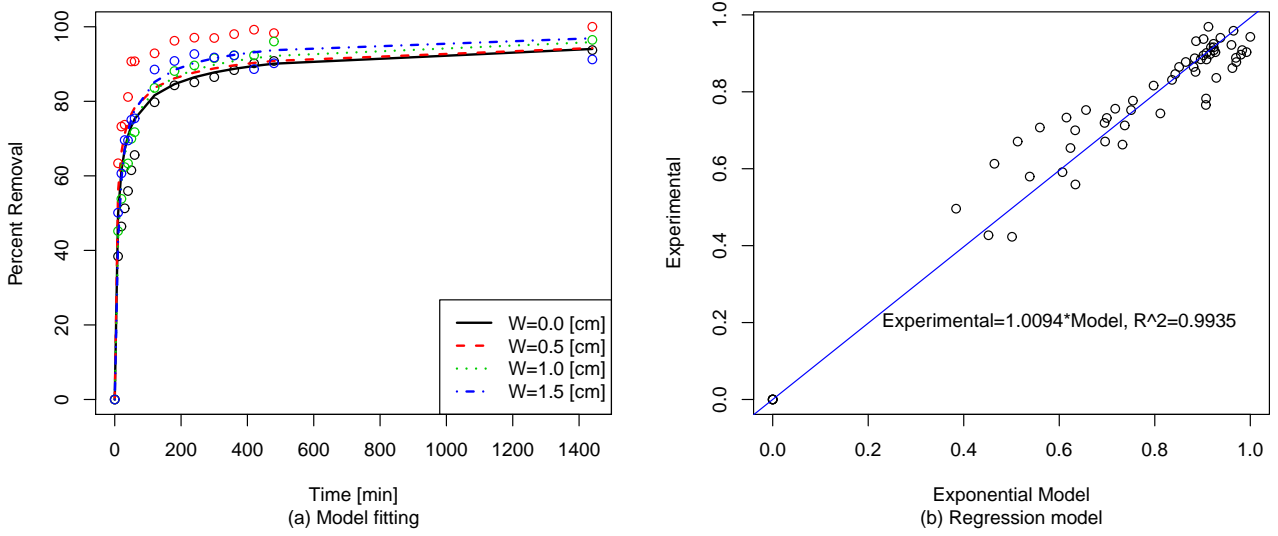}
\caption{Experimental versus exponential model of the removal of $Pb^{2+}$.}
\label{fig_matpb}       
\end{figure}
In Figs.~\ref{fig_matpb} and \ref{fig_mat2}, performance of exponential models are given. As we can see, models are able to fail if percent removal changes with respect thicknesses. They estimate that $0$ $cm$ to $1.5$ $cm$ removal behave either increasing or decreasing. However, $W=0.5$ $cm$ behaves differently in both contaminant removals. Overall, as regression curves show, models are able to estimate better than any multiple linear regression model. 
 
In this paper, Gaussian Process (GP) notation was adopted from \cite{gramacy2020surrogates}. The nonlinear regression alternative, GPs can be simply written as multivariate Gaussian distributions via covariance structure defined over time, dependent variable, or spatial correlations. For instance, $\textbf{X}=(\textbf{T}, \textbf{pH}, \textbf{W})$ with general exponential decaying covariance structure can be expressed as $\sigma^2=\Sigma(X, X)=ve^{-\sum_p{w_p||X_p-X_p||^2}}+\epsilon I$ where $p$ is number explanatory variables (e.g., three for $Pb^{2+}$ and two for Methylene Blue), $v, w_p$ are hyperparameters to be optimized for best describing curve of the process, i.e., percent removal in this paper and $\epsilon I$ is jittering with e.g., $\epsilon=1.490116e^{-08}$ needed for inverse of the covariance matrix. In Eq.~(\ref{eqn_gpfit}), fitting a GP models is given \cite{gramacy2020surrogates,shi2014gpfda}.
\begin{eqnarray}
Given \quad observations \quad (\textbf{X},Y) \nonumber \\
\textit{Calculate} \quad \sigma^2=\Sigma(X,X)=ve^{-\sum_p{w_p||X_p-X_p||^2}}+\epsilon I \nonumber \\ 
\textit{solve as} \quad \Sigma_n^{-1} \nonumber\\
\mu(X)=\Sigma(X,X) \Sigma_n^{-1} Y \nonumber\\
\sigma^2(X)=\Sigma(X,X)-\Sigma(X,X)\Sigma^{-1}\Sigma(X,X)^T
\label{eqn_gpfit}
\end{eqnarray}

With new $X'$ observations with size $m\times p$, GP-based predictions can be calculated by using Eq.~(\ref{eqn_gp}).

\begin{eqnarray}
With \quad new \quad \textbf{X'} \quad obervations \quad (\textbf{X,X'},Y) \nonumber \\
\textit{Calculate} \quad \sigma^2=\Sigma(X,X')=ve^{-\sum_p{w_p||X_p-X'_p||^2}}+\epsilon I \nonumber \\
\sigma^2(X')=\Sigma(X,X')-\Sigma(X,X')\Sigma^{-1}\Sigma(X,X')^T \nonumber\\
Obtain \quad predictions: \quad \Phi^{-1}(m, \mu(X), \sigma^2(X')) 
\label{eqn_gp}
\end{eqnarray}

For $Pb^{2+}$ hyperparameters optimized are $v=0.3852$, $w_1=0.7839$ for time, $w_2=2.8869$ for pH, and $w_3=2.859\times10^{-9}$ for thickness $W$. Similarly for Methylene Blue $v=0.2397$, $w_1=14.6899$ for time and $w_2=2.2309$ for thickness $W$ \cite{shi2014gpfda}.

\begin{figure}[!htb] 
\centering
\includegraphics[scale=.55]{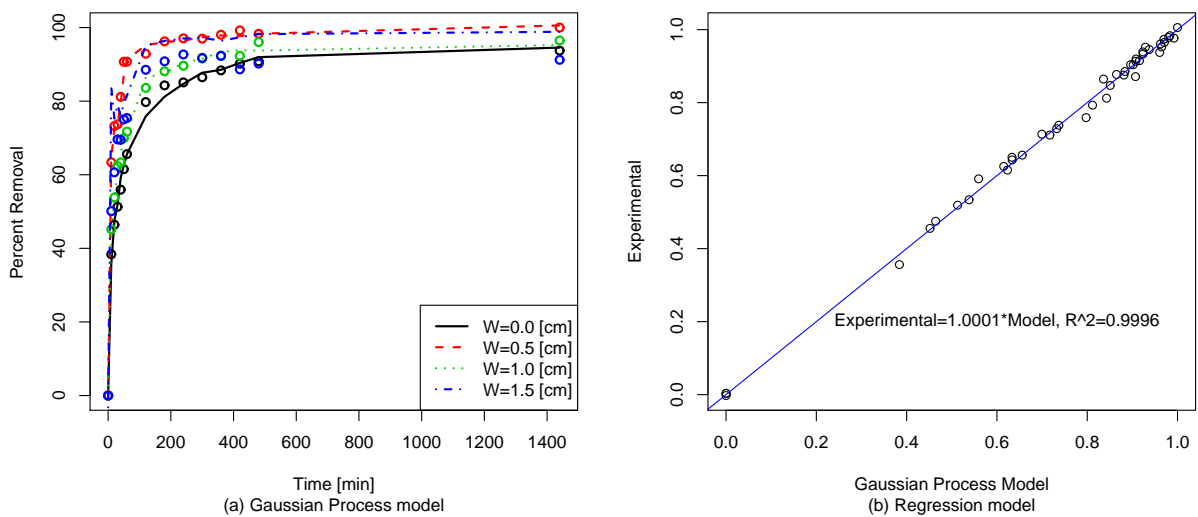}
\caption{Experimental versus Gaussian Process model of the removal of $Pb^{2+}$.}
\label{fig_gppb}       
\end{figure}

\begin{figure}[!htb] 
\centering
\includegraphics[scale=.5]{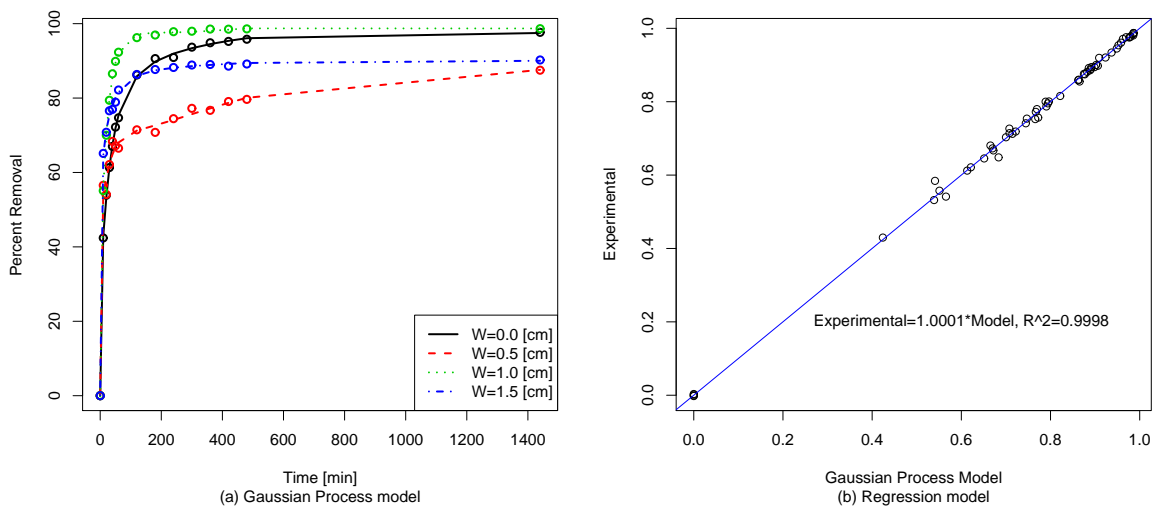}
\caption{Experimental versus Gaussian Process model of the removal of Methylene Blue.}
\label{fig_gpmb}       
\end{figure}

In Figs.~\ref{fig_gppb} and \ref{fig_gpmb}, performance of GP models are given. As we can see, GP models are able to explain the percent removal changes with respect thicknesses with pH ($Pb^{2+}$) and without pH (Methylene Blue). They are able to estimate  the varying $W=0.5$ $cm$ behavior in both contaminant removals. Moreover, GP prediction performance was shown in $Pb^{2+}$ for $W=1.5$ $cm$. From regression curve, it can be seen that GP shows decent performance by following the straight line with coefficient of $1.0001$. 

\begin{figure}[!htb] 
\centering
\includegraphics[scale=.45]{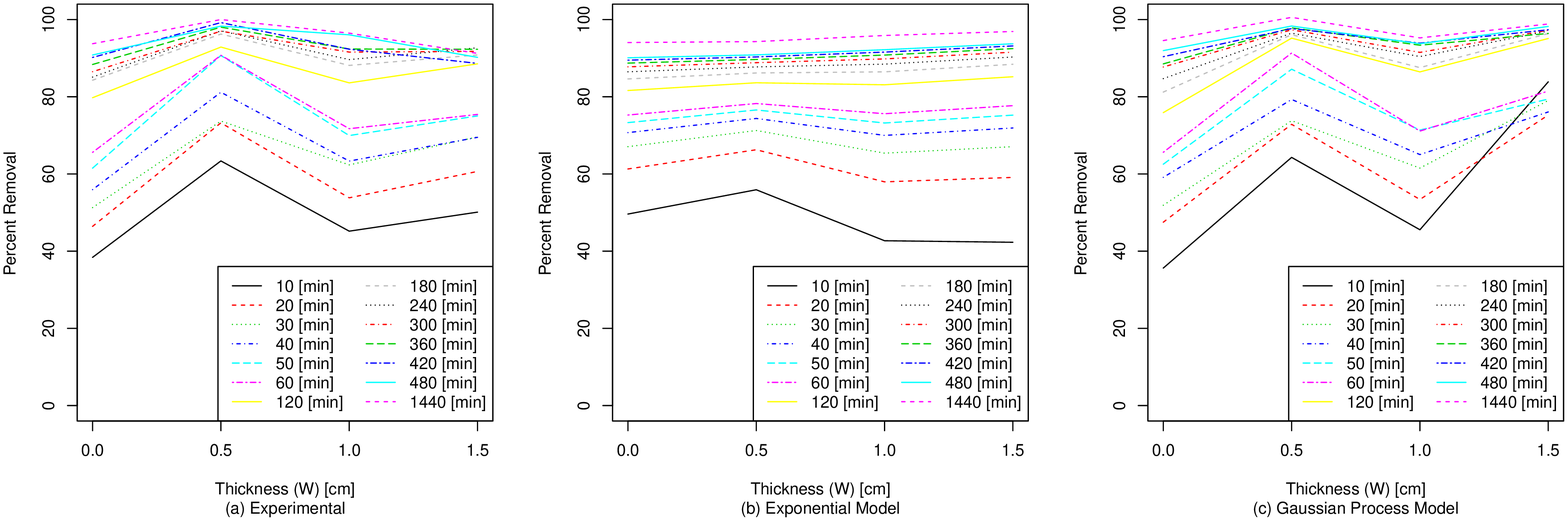}
\caption{Experimental, exponential, and Gaussian Process models for the removal of $Pb^{2+}$.}
\label{fig_matgppb}       
\end{figure}

\begin{figure}[!htb] 
\centering
\includegraphics[scale=.45]{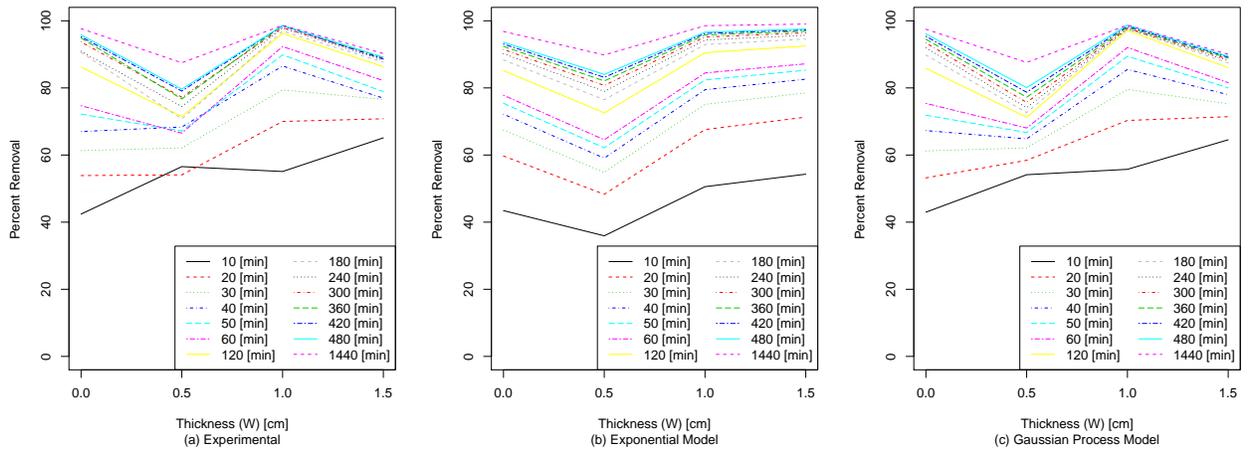}
\caption{Experimental, exponential, and Gaussian Process models for the removal of Methylene Blue.}
\label{fig_matmb}       
\end{figure}

In Figs.~\ref{fig_matgppb} and \ref{fig_matmb}, performance of Exponential and GP models are compared. We can see the behavior of GP estimates are much closer to observed percent removals at different times and thickness levels. We can also see that for $Pb^{2+}$ optimum thickness level is about $W=0.5$ and for  Methylene Blue is about $W=1.0$ $cm$.  
 
\subsection{Conclusion}
\label{sctconc}
This study investigated the possibility of applying pinecone derived adsorbents; PCP and PCBC in PABs for the remediation of $Pb^{2+}$ from a simulated unconfined aquifer. Results have indicated that both PCP and PCBC possess adsorptive potentials capable of removing $Pb^{2+}$ from a contaminated groundwater system. PCBC has depicted higher adsorption capacity both in the first run and the second run with spiked $Pb^{2+}$ concentrations. PCP removed $72.0\%$ and $63.02\%$ of $Pb^{2+}$ in the first and second runs respectively. PCBC on the other hand was capable of removing a total of $86.94\%$ and $82.12\%$ of $Pb^{2+}$ in the first and second run respectively. Due to the major influence of functional groups on the adsorption of $Pb^{2+}$ using PCP, it can deduced that the adsorption mechanism was chemisorption while PCBC exhibited a physi-sorption mechanism. 

Nonlinear regression models were fit to explain the percent removal. Gaussian Process regression performs excellent which can be used to predict percent removal in between thicknesses against time and pH values. Behavior of the removal after certain time can also be predicted at certain thickness.   
 
\section*{Acknowledgments}
This paper is based upon work supported by DOE-EM Grant No. 0000473 and Freeda M. Johnson Laboratory  for Environmental Research, Benedict College. This study is also partially supported by the Center for Connected Multimodal Mobility ($C^{2}M^{2}$) (USDOT Tier 1 University Transportation Center) headquartered at Clemson University, Clemson, South Carolina. Any opinions, findings, and conclusions or recommendations expressed in this material are those of the authors and do not necessarily reflect the views of the Center for Connected Multimodal Mobility ($C^{2}M^{2}$) and the official policy or position of the USDOT/OST-R, or any State or other entity, and the U.S. Government assumes no liability for the contents or use thereof. It is also partially supported by NSF SC EPSCoR/IDeA, 1719501, 1436222, 1954532, and 1400991.
\begin{singlespace}
\bibliographystyle{elsarticle-harv}
\bibliography{transport_bibliography2}
\end{singlespace}






\end{document}